\documentstyle[11pt]{article}
\begin{document}
\title{Relativistic hodograph equation for a two-dimensional
stationary isentropic hydrodynamical motion}
\author{I.M. Khalatnikov$^{1,2}$ and \ A.Yu. Kamenshchik$^{1,3}$}
\date{}
\maketitle
\hspace{-6mm}$^{1}${\em L.D. Landau Institute for
Theoretical Physics, Russian
Academy of Sciences, Kosygin str. 2, Moscow, 119334, Russia}\\
$^{2}${\em Tel Aviv University,
Tel Aviv University, Raymond and Sackler
Faculty of Exact Sciences, School of Physics and Astronomy,
 Ramat Aviv, 69978, Israel}\\
$^{3}${\em Dipartimento di Scienze Fisiche e Matematiche,
Universit\`{a} dell'Insubria, via Valleggio 11, 22100 Como,
Italy}
\begin{abstract}
We derive a relativistic hodograph equation for a two-dimensional
stationary isentropic hydrodynamical motion. For the case of stiff
matter, when the velocity of sound coincides with the light speed,
the singularity in this equation disappears and the solutions
become regular in all hodograph plane.
\end{abstract}
The non-relativistic problem for two-dimensional stationary
isentropic hydrodynamical motion was formulated and solved by
Chaplygin in 1902 (see \cite{LL}). The solution was based on a
transformation from the physical (coordinate) plane $x_1,x_2$ to
the velocity plane $v_1,v_2$. This transformation is called
hodograph transformation. The beauty of this Chaplygin's approach
consists in the fact that it allows to undertake a transition from
two non-linear hydrodynamical equations to one linear partial
derivatives equation for a potential in the hodograph plane.

Below we formulate and solve general stationary two-dimensional problem
for the relativistic hydrodynamics. We use some results from the work by
Khalatnikov (1954) \cite{Kh}, where the general approach to the
relativistic hydrodynamics was formulated and the general one-dimensional
non-stationary Chaplygin problem was solved.

To begin with let us introduce the variables: the relativistic
velocity $u_i$ such that\footnote{We choose the velocity of light
equal to zero.}
\begin{equation}
u_i u^i = u_1^2 + u_2^2 - u_4^2 = -1,
\label{vel}
\end{equation}
the enthalpy per particle $w$ and the particle density $n$ \cite{LL,Kh}.
From the quasi-potentiality condition \cite{Kh} in the relativistic
hydrodynamics
\begin{equation}
w u_i = \frac{\partial \varphi}{\partial x^i},
\label{quasi}
\end{equation}
follows that for the potential $\varphi$:
\begin{equation}
d\varphi = wu_1 dx_1 + wu_2 dx_2.\label{quasi1}
\end{equation}
Let us make a Legendre transformation to the hodograph plane,
going from the variables $x_1$ and $x_2$ to the variables $u_1$
and $u_2$, and introducing the potential
\begin{equation}
\chi = \varphi -x_1 w u_1 - x_2 w u_2.
\label{Legendre}
\end{equation}
For the potential motion we write instead of the relativistic
Euler equation its first integral, i.e. the Bernoilli equation
\cite{Taub,Kh}
\begin{equation}
wu_4 = const.
\label{Bern}
\end{equation}

In the hodograph plane we introduce the ``angles'' $\eta$ and $\theta$
such that
\begin{eqnarray}
&&u_1 = \sinh\eta \cos\theta,\nonumber \\
&&u_2 = \sinh\eta \sin\theta,\nonumber \\
&&u_3 = \cosh\eta.
\label{angles}
\end{eqnarray}
In these variables the differential of the potential $\chi$ is
\begin{eqnarray}
&&d\chi = -w[(x_1\cos\theta + x_2\sin\theta)\cosh\eta d\eta \nonumber\\
&&+(-x_1 \sin\theta +x_2\cos\theta_)\sinh\eta d\theta]\nonumber\\
&&-(x_1\cos\theta + x_2\sin\theta)\sinh\eta dw.
\label{dif}
\end{eqnarray}
Taking into account the Bernoilli relation (\ref{Bern}) and omitting the
normalization factor we have
\begin{eqnarray}
&&\frac{\partial \chi}{\partial \eta} = -\frac{1}{\cosh^2\eta}
(x_1 \cos\theta + x_2 \sin\theta),\nonumber \\
&&\frac{\partial \chi}{\partial \theta}
=-\tanh\eta(-x_1\sin\theta+x_2\cos\theta). \label{deriv}
\end{eqnarray}
The absolute value of the spatial velocity $v$ is
\begin{equation}
v = \tanh\eta.
\label{spatial}
\end{equation}

The inverse transformation to the physical space is
\begin{eqnarray}
&&x_1 = -\left(\cos\theta\frac{\partial \chi}{\partial v} -
\frac{\sin\theta}{v}\frac{\partial \chi}{\partial \theta}\right),
\nonumber \\
&&x_2 = -\left(\sin\theta\frac{\partial \chi}{\partial v} +
\frac{\cos\theta}{v}\frac{\partial \chi}{\partial \theta}\right).
\label{inverse}
\end{eqnarray}
The relation between potentials $\varphi$ and $\chi$ is
\begin{equation}
\varphi = \chi - v\frac{\partial \chi}{\partial v}.
\label{pot-pot}
\end{equation}
Note that these relations have the same form as their non-relativistic
analogs \cite{LL}.

Now, we can deduce the equation for the potential $\chi$ using the
 continuity equation
\begin{equation}
\frac{\partial}{\partial x_i}(n u^i) =
\frac{\partial}{\partial x_1}(n u_1) +
\frac{\partial}{\partial x_2}(n u_2) = 0,
\label{cont}
\end{equation}
or, in terms of the variables $w,v$ and $\theta$
\begin{equation}
\frac{\partial}{\partial x_1}\left(\frac{n}{w} v\cos\theta\right) +
\frac{\partial}{\partial x_2}\left(\frac{n}{w} v\sin\theta\right) = 0.
\label{cont1}
\end{equation}
To get the equation for $\chi$, we should meke in the preceding
equation (\ref{cont1}) the transition to the variables $v$ and
$\theta$:
\begin{equation}
\frac{\partial^2\chi}{\partial\theta^2} +
\frac{v^2(1-v^2)}{1-\frac{v^2}{c^2}}\frac{\partial^2\chi}{\partial
v^2} + v\frac{\partial\chi}{\partial v} = 0, \label{main}
\end{equation}
where the sound velocity $c$ is defined as \cite{Kh}
\begin{equation}
c^2 = \frac{n}{w} \frac{\partial w}{\partial n}.
\label{sound}
\end{equation}

The relativistic hodograph equation (\ref{main}) together with
Eqs. (\ref{inverse})
 play a role of equations of motion. Thus, the problem of solution
of nonlinear equations of motion is reduced to the solution of the
linear equation for $\chi$ in the hodograph plane. However, the
boundary conditions for Eq. (\ref{main}) are nonlinear (for
details see \cite{LL}).

Let us pay a special attention to the crucial difference with respect
to the non-relativistic case \cite{LL}
\begin{equation}
\frac{\partial^2\chi}{\partial\theta^2} +
\frac{v^2}{1-\frac{v^2}{c^2}}\frac{\partial^2\chi}{\partial v^2}
+ v\frac{\partial\chi}{\partial v} = 0,
\label{main1}
\end{equation}
which consists in the appearance of the factor $(1-v^2)$ in front of
the term $\frac{\partial^2\chi}{\partial v^2}$. Both these equations
have a singularity at $v=c$ which corresponds a transition to the
supersonic regime. The supersonic regime brings also another problem,
which is connected with the possible vanishing of the Jacobian
\begin{equation}
\frac{\partial(x_1,x_2)}{\partial(\theta,v)} = \frac{1}{v}
\left[\left(\frac{\partial^2\chi}{\partial v \partial \theta}
-\frac{1}{v}\frac{\partial \chi}{\partial\theta}\right)^2
+\frac{v^2(1-v^2)}{1-\frac{v^2}{c^2}}\left(\frac{\partial^2\chi}
{\partial v^2}\right)^2\right],
\label{Jacobi}
\end{equation}
which in the subsonic regime is always positive. Nullification of
this Jacobian on some ``limiting'' line $v=v(\theta)$ in the
supersonic regime makes the velocity $v$ complex on one side of
this limiting line
 \cite{LL}. It signifies that the appearance of shock waves is unavoidable
in this regime.

It is extremely interesting that these problems are absent for the
relativistic motion of a fluid with the stiff equation of state
when  pressure coincides with  energy density and the sound
velocity coincides with the speed of light ($c=1$). In this case the
factors $(1-v^2)$ and $\left(1-\frac{v^2}{c^2}\right)$ cancel each
other. Naturally, the supersonic regime seems to be impossible for
the stiff fluid, because such a regime would be also a
superluminal one. Note, however, that even considering
supersonic/superluminal regime, we always have real solutions of
the Chaplygin equation and the potential and velocity cannot
become complex.

The consideration presented above can be treated as an
introduction to the further study of two-dimensional motion in the
relativistic hydrodynamics, which could have astrophysical and
cosmological applications.

Authors are grateful to R.A. Syunyaev, who has pointed out to an
opportunities to apply the described methods to the study of some
astrophysical problems.

\section*{Appendix.Some general remarks concerning the relativistic
hydrodynamics} One-dimensional relativistic non-stationary
Chaplygin problem was studied in paper by Khalatnikov \cite{Kh}.
This study was undertaken for further development of the
Landau-Fermi theory of high-energy multiple particle production.
Here, we would like to mention only its main results and write
down the relativistic non-stationary one-dimensional Chaplygin
equation.

Starting with the quasi-potentiality condition (\ref{quasi}) and
introducing the velocity components as
\begin{equation}
u_1 = \sinh\eta,\ \ u_4 = \cosh\eta,
\end{equation}
we undertake the Legendre transformation
\begin{equation}
\chi = \varphi -w u_1 x_1 -w u_4 x_4.
\end{equation}
Acting as above we come to the equation
\begin{equation}
\frac{1}{n}\frac{\partial n}{\partial w} \left(\frac{\partial
\chi} {\partial w} -
\frac{1}{w}\frac{\partial^2\chi}{\partial\eta^2}\right) +
\frac{\partial^2\chi}{\partial w^2} = 0.
\end{equation}
Using the expression (\ref{sound}) for the sound velocity and
introducing a new variable $y=\ln w$, one gets the following equation
\begin{equation}
c^2\frac{\partial^2\chi}{\partial y^2} + (1-c^2)\frac{\partial \chi}
{\partial y} - \frac{\partial^2\chi}{\partial \eta^2} = 0.
\label{wave}
\end{equation}
This equation represents a solution of the one-dimensional
non-stationary problem in relativistic hydrodynamics.
As before it should be accompanied by the choice
of the special boundary conditions and by formulas, which
express the time and spatial coordinate variable as functions
of the potential $\chi$:
\begin{eqnarray}
&&t = e^{-y}\left(\frac{\partial\chi}{\partial y}\cosh\eta -
\frac{\partial\chi}{\partial\eta}\sinh\eta\right),\nonumber \\
&&x_1 = e^{-y}\left(\frac{\partial\chi}{\partial y}\sinh\eta -
\frac{\partial\chi}{\partial\eta}\cosh\eta\right).
\end{eqnarray}

Again, in the
stiff matter case $c = 1$, Eq. (\ref{wave}) simplifies essentially and
becomes a simple one-dimensional wave equation.

\end{document}